\documentclass[12pt]{article}
\usepackage[margin=0.8 in]{geometry}    
\usepackage[title,titletoc,toc]{appendix}
\usepackage{authblk}
\usepackage{hyperref}     
 \usepackage{color}
 \usepackage[most]{tcolorbox}
 \tcbset{colback=yellow!10!white, colframe=red!50!black, 
        highlight math style= {enhanced, 
            colframe=red,colback=red!10!white,boxsep=0pt}
        }        
\usepackage{graphicx}
\usepackage{amssymb}
\usepackage{amsmath}
\usepackage{amsfonts}
\usepackage{epstopdf}
\newcommand{\olra}{\overleftrightarrow}

\newcommand{\p}{\partial}
\newcommand{\DDx}[2]{\frac{\dd #1}{\dd #2}}
\newcommand{\ddx}[2]{\frac{d #1}{d #2}}
\newcommand{\ppx}[2]{\frac{\p #1}{\p #2}}

\newcommand{\ppy}{\frac{\partial }{\partial y}}

\newcommand{\ppys}{\frac{\partial ^2}{\partial y^2}}

\newcommand{\lo}{\Lambda_0}
\newcommand{\lm}{\Lambda}
\newcommand{\hf}{{1\over 2}}
\newcommand{\be}{\begin{equation}}
\newcommand{\br}{\begin{eqnarray}}
\newcommand{\er}{\end{eqnarray}}
\newcommand{\ee}{\end{equation}}
\newcommand{\bt}{\begin{tabular}}
\newcommand{\et}{\end{tabular}}
\newcommand{\bc}{\begin{tcolorbox}}
\newcommand{\ec}{\end{tcolorbox}}
\newcommand{\bp}{\bar p}

\newcommand{\dd}{\delta}
\newcommand{\DD}{\Delta}

\newcommand{\CD}{{\cal D}}
\newcommand{\Dp}{\frac{d^Dp}{(2\pi)^D}}

\newcommand{\eps}{\epsilon}

\newcommand{\si}{\sigma}

\newcommand{\Dt}{\frac{D}{2}}
\numberwithin{equation}{section}
\title{Mapping from Exact RG to Holographic RG in Flat Space}
\author[1,2,3]{B. Sathiapalan \thanks{\href{mailto:bala@imsc.res.in}{bala@imsc.res.in}}}
\affil[1]{Institute of Mathematical Sciences, CIT Campus, Tharamani, Chennai 600113, India}
\affil[2]{Homi Bhabha National Institute\\Training School Complex, Anushakti Nagar, Mumbai 400085, India}
\affil[3]{Chennai Mathematical Institute, H1, SIPCOT IT Park\\ Siruseri
Kelambakkam 603103,
India}

\begin{document}

\hspace{12cm} IMSc/2024/03
{\let\newpage\relax\maketitle}
\begin{abstract}
In earlier papers a method was given for constructing from first principles a holographic bulk dual action in Euclidean AdS space for a Euclidean CFT on the boundary. The starting point was an Exact RG for the boundary theory. The bulk action is obtained from the evolution operator for this ERG followed by a field redefinition. This procedure guarantees that the boundary correlators are all recovered correctly. In this paper we  use the same method in an attempt to construct a holographic dual action for the free $O(N)$ model where the bulk is  flat Euclidean space with a plane boundary wall. The scalar cubic interaction is found to be local (in $D=3$) but depends on the distance from the boundary - which can be interpreted as a non constant background dilaton field. The spin 2 - scalar -   scalar interaction is found to be non local - in contrast to the AdS case. A field redefinition that makes the kinetic term quartic in derivatives can be done to  eliminate this non locality. It is shown that the action can be obtained by gauge fixing an action that has the linearized gauge invariance associated with general coordinate invariance. Boundary correlators (two point and three point) are shown to be reproduced by bulk calculations - as expected in this approach to holography.

\end{abstract}

\newpage 
\tableofcontents 
\newpage
\section{Introduction}

The  AdS/CFT correspondence \cite{Maldacena,Polyakov,Witten1,Witten2}   is remarkable because it relates a quantum gravity theory to a flat space field theory in one lower dimension. While there is ample evidence that it is true in a number of cases, a proper understanding of why such a correspondence should exist is missing. 

The idea of holographic RG provided some insight into this correspondence
[\cite{Akhmedov}-\cite{SSLee:2012}]. 
 A suggestion for deriving holographic RG from Exact RG [\cite{Wilson}-\cite{Wilson2}] was was made in \cite{Sathiapalan:2017,Sathiapalan:2019}. 

In \cite{Sathiapalan:2017,Sathiapalan:2019} a method was given for constructing a $D+1$ dimensional holographic dual action in Euclidean space starting from an Exact RG equation (in the form given by Polchinski \cite{Polchinski})  for a $D$-dimensional boundary theory of a free scalar field. The evolution operator for this ERG is a functional integral with a non local $D+1$ dimensional action. A field redefinition maps this to a local action of a scalar field in Euclidean $AdS_{D+1}$ and thus explains the correspondence for this special situation.
 This construction has been generalized to composite (``single trace") operators in the $O(N)$ model, such as the composite scalar, vector current and energy momentum tensor \cite{Sathiapalan:2020,Dharanipragada:2022}. The boundary theory is that of scalar, gauge field, and graviton respectively. In \cite{Sathiapalan:2020} the cubic interaction for the scalar composite was worked out and found, very surprisingly,  to be local at the UV cutoff scale (eg Planck length scale) rather than the much larger $AdS$ radius scale as one might naively expect from an RG. This is also true for the scalar-scalar-graviton interaction term \cite{Dharanipragada:2023}. This method thus potentially gives a construction from first principles of a holographic bulk dual. The bulk dual has gravity as one of the modes so it is a quantum gravity theory.  The existence of this dual follows from the exact RG equation albeit in the large $N$ approximation. This then can be viewed as the first steps towards a derivation of the AdS/CFT correspondence for this very special case
 and provides a different way of thinking about the correspondence.
The fact that the bulk theory is local is not obvious from this construction and it is also not clear whether it is true in general.

A natural question that arises is - what is special about AdS space?
Given that the bulk theory has quantum gravity one would morally expect that the bulk metric should not be fixed to be pure AdS. In which case holographic dualities should exist for other spaces too. Of course if the boundary is at infinity 
it does make sense  to specify and fix the asymptotic behaviour of the metric. So one might argue that the only  asymptotic behaviour that has holographic properties is an asymptotically AdS space.   In the boundary this implies a conformal field theory - a field theory at its fixed point. 

Now a fixed point theory can have a {\em finite} UV cutoff. It is obtained as a solution to an exact RG (ERG) fixed point equation \footnote{See \cite{Dutta:2020} for a concrete example.}. In the AdS bulk this would correspond to a boundary that is at a finite distance rather than at asymptotic infinity. This is like quantum gravity in a box. But in a finite box the notion of asymptotic behaviour is ambiguous. So this suggests that other spaces might also have holographic properties.

So then one of the obvious questions that this construction raises is whether maps to  spaces other than AdS is possible in the ERG approach. This was partially answered in \cite{Dharanipragada:2022} where an analogous map was made to de Sitter space - after an analytic continuation of the Euclidean ERG ``time". This also made contact with earlier work on dS/CFT correspondence[\cite{Witten:2001}-\cite{Larsen:2002}]. Flat space times are also of obvious interest and one can ask whether this approach can give some insight into flat space holography which has been discussed a lot in the literature \cite{Strominger:2017,Bagchi:2016,Gupta:2020,Jain:2023}. In this paper we construct a map that takes the ERG evolution operator functional integral to an action in  flat space. As before the kinetic terms are local . The cubic term in the action is, surprisingly enough,  local for the scalar self interaction for the case $D=3$. However it depends on the distance from the boundary. But the  graviton scalar interaction is non local. This is in contrast to the situation in $AdS$ where the same procedure gives local interactions in the above cases. One also finds that a field redefinition mitigates the non locality of this interaction at the price of obtaining a kinetic term that is quartic in derivatives. This is reminiscent of conformal gravity.

The other important question is whether this bulk flat space theory can be obtained by gauge fixing a general coordinate invariant theory.  A gauge invariant coupling of the spin 2 field to scalars is given that reduces on gauge fixing to the interaction obtained using the ERG approach. To make the cubic scalar self coupling  coordinate invariant one has to invoke a dilaton coupling with a non constant expectation value for the dilaton field. Thus with this assumption the bulk theory is consistent with general coordinate invariance.

The ERG approach always gives a holographic bulk action that reproduces the correlators of the boundary theory. It can be mapped to actions in AdS or flat space. A relevant question then is whether the bulk theory obtained is ``interesting".  One property that makes a theory interesting is locality. The AdS case seems to have this locality property for the cubic interactions. The flat space example constructed  here is not local.  But it is instructive - it sheds some light on the question of why AdS spaces are special. Furthermore, being dual to free theory on the boundary, it may still be a well defined bulk theory of an interacting massless spin 2 particle. 

 It is not clear whether the correspondence derived here has any connection with earlier approaches to this problem of flat space holography.  For one, we are in Euclidean space. The Minkowski structure was crucial in all the earlier works. Whether analytic continuation, analogous to what happens in the dS/CFT correspondence, can be fruitful here is a question for the future.

This paper is organized as follows. In Section \ref{2} we give the zero dimensional case following \cite{Dharanipragada:2022}. The map takes an ERG equation to a Euclidean harmonic oscillator. Since there are no momenta or cutoff here, the connection to RG is formal.  In Section \ref{3} we give the field theory version. In Section \ref{4} we show how the two point function of the boundary theory is reproduced from the bulk. The application of this to $O(N)$ models is immediate and is given in Section \ref{5} where some cubic interactions and correlators are evaluated. Gauge invariance is also discussed in this section. Section \ref{6} gives a summary and conclusion.

\section{Zero Dimensions} \label{2}

For mathematical simplicity we first discuss the Polchinski ERG equation for a zero dimensional theory. 
Here the boundary is zero dimensional. When the Euclidean RG time is added it becomes a one dimensional theory and an imaginary time Schroedinger equation (i.e. a diffusion equation) is obtained. There is no momentum and hence no cutoff in the boundary theory. So the connection with ERG equation is purely formal. In the next section we will see that the results are easily generalized to $D$ dimensions.

We remind the reader of the form of Polchinski's ERG equation for a $D$ dimensional field theory:
If we write a  scalar field Wilson action as
\[
S[\phi]= \hf \int \Dp \phi(p)G(p,\tau)^{-1}\phi(-p) +S_{I,\lm}[\phi]
\]
where $G(p)$ is the regulated low energy propagator and $S_{I,\lm}[\phi]$ is the interacting part of the action, then Polchinski's ERG equation is

\be   \label{polch0}
\frac{\p}{\p \tau} \Psi[\phi(p),\tau] = -\hf \int\Dp \dot G(p)\frac{\dd^2 \Psi[\phi(p),\tau]}{\dd \phi(p)\dd \phi(-p)}
\ee
where $\Psi = e^{-S_{I,\lm}[\phi]}$. $G(p,\tau) $ for a free scalar field could be of the form ($\lm \approx e^{-\tau}$)  
\[
G(p,\lm)= \frac{e^{-p^2/\lm^2}}{p^2}
\]
But we allow more general possibilities. The power of $p$ in the denominator is left free (for composite fields). The cutoff function is also left free.

For the zero dimensional case $D=0$, - which is like the heat equation of a point particle - $\phi (p)$ is just one real function - and we use the symbol $x$ for it (see below). This is also like the Schroedinger equation in imaginary time (and a time dependent mass).

\subsection{ERG Equation}\label{ESE}
We reproduce a calculation from \cite{Dharanipragada:2022} where the ERG equation in zero dimensions is mapped to a harmonic oscillator.
The Euclidean ERG equation is
\be  \label{C5}
\frac{\p \Psi(x,\tau)}{\p \tau}= -\hf \dot G(\tau)\frac{\p^2 \Psi(x,\tau)}{\p x^2}
\ee

As mentioned above, this is of the form of a Polchinski ERG equation   for 
${\cal S}$ defined by $\Psi\equiv e^{-{\cal S}}$.
We change variables and define $x=f(\tau)y$ with $f^2= -\dot G$ and $\Psi = e^{-\hf \alpha y^2}\bar \Psi$, to find
\be   \label{C5.5}
\frac{\p \bar\Psi}{\p \tau}= (\hf  \dot \alpha -\alpha \frac{\dot f}{f} +\hf \alpha^2)y^2\bar \Psi +(\frac{\dot f}{f} y\ppy - \alpha y\ppy) \bar \Psi -\hf \alpha \Psi + \frac{1}{2}\ppys \bar \Psi
\ee

We impose the condition
$\alpha = \frac{\dot f}{f}$ and the equation becomes

\be   \label{C6}
\frac{\p \bar \Psi}{\p \tau}=\frac{1}{2} [-\underbrace{e^{\ln f}(\frac{d^2}{d\tau^2} e^{-\ln f})}_{=~\omega_0^2}y^2+\ppys-\alpha]\bar \Psi
\ee
If we set the combination indicated by brackets to equal a constant, $\omega_0^2$,
one obtains a (Euclidean) harmonic oscillator equation  \footnote{This equation can be interpreted as an equation for the canonical density matrix of a harmonic oscillator in thermal  equilibrium with inverse temperature $\tau$.}
:
\be   \label{C7}
\frac{\p \bar \Psi}{\p \tau}=\frac{1}{2} [\ppys-\omega_0^2y^2-\alpha]\bar \Psi
\ee

The term $\alpha$  (which is $\frac{\dot f}{f})$ in \eqref{C6} provides a multiplicative scaling
$e^{-\hf \int_{\tau_i}^\tau d\tau '~\p_{\tau'}\ln f}=(\frac{f(\tau_i)}{f(\tau)})^\hf$ of $\bar \Psi$  and has no effect on the physics.
And $f$ obeys
\be
\frac{d^2}{d\tau^2}\frac{1}{f}= \omega_0^2 \frac{1}{f}
\ee

\subsection{Functional formalism}\label{ESF}

We now turn to the evolution operator for the ERG equation of the last section. This equation looks like an imaginary time Schroedinger equation for a free particle with a time dependent mass $M_E(\tau)=\frac{1}{\dot G(\tau)}$. The evolution operators is the following  Euclidean functional integral:
\be  \label{B1}
\Psi(x_,\tau) =\int dx_i \int_{\begin{array}{ccc}
x(\tau_i)&=&x_i\\
x(\tau)&=&x
\end{array}} {\cal D}x ~e^{\hf \int _{\tau_i}^\tau \frac{1}{\dot G(\tau')} \dot x^2 d\tau'} \Psi(x_i,\tau_i)
\ee

As before, let $x(\tau)=f(\tau)y(\tau)$ with $f^2(\tau) = -\dot G(\tau)$.
Substitute this in \eqref{B1}.
\[
S=\hf \int d\tau~ ( \dot y^2+ (\frac{\dot f}{f})^2 y^2 + 2 \frac{\dot f}{ f} \dot y y)
\]
\[
=\hf \int d\tau~ [ \dot y^2+ (\frac{d \ln f}{d\tau})^2 y^2 -(\frac{d^2}{d\tau^2} \ln f ) y^2] +\hf\int d\tau~\frac{d}{d\tau}(\frac{d \ln f}{d\tau}y^2)
\]
Thus, upto the boundary term, the action is
\be
S=
\hf \int d\tau~ [ \dot y^2+  e^{\ln f}(\frac{d^2}{d\tau^2} e^{-\ln f }) y^2] 
\ee
Now choose
\be   \label{B4.5}
e^{\ln f}(\frac{d^2}{d\tau^2} e^{-\ln f })=\omega_0^2
\ee
and we get

\be \label{B3.6}
\bar S=
\hf \int d\tau~ [ \dot y^2 + \omega_0^2 y^2] 
\ee

which is the  action for a Euclidean harmonic oscillator - or a Hamiltonian for a harmonic oscillator \footnote{Related results were obtained in \cite{Padmanabhan:2017,Ramos}. These maps can be understood as canonical transformations in Quantum Mechanics \cite{Anderson:1992}.}. And we define $\bar \Psi$ by absorbing the contribution from the boundary term:

\be 
\underbrace{e^{-\hf  \frac{d\ln f(\tau)}{d\tau}y^2(\tau)} \Psi(f(\tau)y,\tau)}_{\bar \Psi(y,\tau)} =\int dy_i \int_{\begin{array}{ccc}
y(\tau_i)&=&y_i\\
y(\tau)&=&y
\end{array}} {\cal D}y ~e^{\hf \int _{\tau_i}^\tau [\dot y^2 +\omega_0 ^2 y^2]d\tau'} \underbrace{e^{-\hf  \frac{d\ln f(\tau_i)}{d\tau}y^2(\tau_i)}\Psi(f(\tau_i)y_i,\tau_i)}_{\bar \Psi(y_i,\tau_i)}
\ee

 As mentioned in the last section,  $\Psi$ can be taken to be $e^{-{\cal S}[x_i,\tau_i]}$ where ${\cal S}$ is a Hamiltonian or Euclideanized Wilson action. Alternatively (depending on what $G(\tau)$ is) it can also be $e^{W[J]}$ - a generating functional or partition function.

\be  \label{B4.7}
\bar \Psi(y,\tau) =\int dy_i \int_{\begin{array}{ccc}
y(\tau_i)&=&y_i\\
y(\tau)&=&y
\end{array}} {\cal D}y ~e^{-\hf \int _{\tau_i}^\tau [\dot y^2 +\omega_0 ^2 y^2]d\tau'} \bar\Psi(y_i,\tau_i)
\ee

The solutions to \eqref{B4.5} written as
\be   \label{B5}
\frac{d^2}{d\tau^2}\frac{1}{f} = \omega_0^2 \frac{1}{f}
\ee

 are of the form
\be
f= A ~sech ~\omega_0(\tau-\tau_0)
\ee
which means $M_E(\tau)=-\frac{1}{\dot G}= \frac{1}{A^2} cosh^2\omega_0(\tau-\tau_0)$ and $G(\tau)$ can be taken to be
\be
G(\tau)= -\frac{A^2}{\omega_0} (tanh ~\omega_0 (\tau-\tau_0) -1)
\ee

\eqref{B4.7} has a $\tau$ independent action. In this case there are well known physical interpretations for the Euclidean theory. The evolution operator, $K(y,\tau;y_i,0)$, where
\be
K(y,\tau;y_i,0)=\int_{\begin{array}{ccc}
y(0)&=&y_i\\
y(\tau)&=&y
\end{array}} {\cal D}y ~e^{-\hf \int _{0}^\tau [\dot y^2 +\omega_0 ^2 y^2]d\tau'}
\ee
is the density operator of a QM harmonic oscillator in equilibrium at temperature specified by $\beta = \tau$.

\section{D dimensional Field Theory} \label{3}

It is a simple matter to generalize the above calculation to the $D$-dimensional case. The first cosmetic step is to go back to field theory notation:  replace $x$ of the boundary field theory by $\phi$ and then make it a function of $p$, the $D$-dimensional Euclidean momentum. We write again below Polchinski's ERG equation: If we write
\[
S[\phi]= \hf \int \Dp \phi(p)G(p,\tau)^{-1}\phi(-p) +S_{I,\lm}[\phi]
\]
where $G(p)$ is the regulated low energy propagator and $S_{I,\lm}[\phi]$ is the interacting part of the action, then Polchinski's ERG equation is

\be   \label{polch1}
\frac{\p}{\p \tau} \Psi = -\hf \int\Dp \dot G(p)\frac{\dd^2 \Psi}{\dd \phi(p)\dd \phi(-p)}
\ee
where $\Psi = e^{-S_{I,\lm}[\phi]}$

The evolution operator that replaces \eqref{B1} is:
\be    \label{evofunda}
\Psi[\phi(p),\tau]=\int{\cal D}\phi_i(p) \int_{\begin{array}{ccc}
\phi(p,\tau_i)&=&\phi_i(p)\\
\phi(p,\tau)&=&\phi(p)
\end{array}}{\cal D}\phi(p,\tau)e^{-\hf\int_{\tau_i}^{\tau}  d\tau'\int \Dp \frac{1}{\dot G(p,\tau')}\dot \phi(p,\tau')\dot \phi(-p,\tau')}\Psi[\phi_i(p),\tau_i]
\ee

Then in the bulk functional  let $\phi(p,\tau)= f(p,\tau) y(p,\tau)$. There is an abuse of notation here - $f(|\vec p|)$, whereas $\phi(\vec p),y(\vec p)$.

One obtains
\be
S=
\hf \int d\tau\int \Dp~ [ \dot y(p,\tau)\dot y(-p,\tau)+  e^{\ln f(p,\tau)}(\frac{d^2}{d\tau^2} e^{-\ln f(p,\tau) }) y(p,\tau)y(-p,\tau)] 
\ee

Now we choose
\be
\frac{d^2}{d\tau^2}\frac{1}{f}=(p^2+m^2) \frac{1}{f}
\ee to obtain

\be   \label{bulkS}
S=
\hf \int d\tau\int \Dp~ [ \dot y(p,\tau)\dot y(-p,\tau)+  (p^2+m^2) y(p,\tau)y(-p,\tau)] 
\ee
giving a $D+1$ dimensional free field theory in flat Euclidean space as promised.
The boundary term will be absorbed as in the zero dimensional case so that
\be
\bar \Psi[y(p,\tau_i),\tau_i]=\bar \Psi[y_i(p),\tau_i]=e^{-\hf \int_p \frac{\p \ln f(p,\tau_i)}{\p\tau} y_i(p)y_i(-p)} \Psi[f(p,\tau_i)y_i(p),\tau_i]
\ee
We have set $y(p,\tau_i)=y_i(p)$. An analogous redefinition is there at $\tau_f$.

Define $\bp^2=p^2+m^2$. The function $f$ has the form
\be  \label{f1}
f(p)= A(p,m) ~sech ~\bp(\tau-\tau_0)
\ee
which means $\dot G(p,\tau)=- A^2 sech^2\bp(\tau-\tau_0)$ and so $G(p,\tau)$ can be taken to be
\be
G(p,\tau)= -\frac{A(p)^2}{\bp} (tanh ~\bp (\tau-\tau_0) -1)
\ee

In order to interpret \eqref{C5} as an ERG equation $G$ should represent a cutoff propagator. This is possible if we interpret $\tau-\tau_0=\frac{1}{\lm}$. Let us take $\tau_0=0$ to begin with. 
\be  \label{G1}
G(p,\lm)= -\frac{A(p)^2}{\bp} (tanh(\frac{\bp}{\lm})-1)
\ee

Then $\tau=0$ corresponds to $\lm =\infty$ and 
\be   \label{g1}
G(p,0) = \frac{A(p)^2}{\bp}
\ee

Choose 
\be  \label{A}
A(p)= \bp^\hf/p
\ee to get
\be
G(p,\lm)=\frac{K(p)}{p^2}= -\frac{1}{p^2} (tanh(\frac{\sqrt {p^2+m^2}}{\lm})-1)
\ee
which is a cutoff low energy propagator.  

When $\lm =\infty$ we have
\be
G(p,0)=\frac{1}{p^2}
\ee
as required.
When $\lm=0$ we have $\tau = \infty$ and  $G(p,\infty)=0$ also as required for a low energy propagator with a UV  cutoff $\lm$.

 The ERG equation \eqref{C5} reads :
\be \label{SERG}
\frac{\p \Psi}{\p \tau}=-\lm^2\frac{\p  \Psi}{\p \lm} =-\hf \int_p\dot G(p,\lm) \frac{\dd^2 \Psi}{\dd \phi(p) \dd \phi(-p)}
\ee
where $\dot G =\frac{\p G}{\p \tau}= -\lm^2\frac{\p G}{\p \lm}$.

Thus to conclude, the action \eqref{bulkS} describes a scalar field in flat space. The ERG evolution operator for the boundary theory is:
\be
K(y_f,\tau_f;y_i,\tau_i)=\int_{y(\tau_i)=y_i,y(\tau_f)=y_f} \CD y~ e^{-\hf \int_{\tau_i}^{\tau_f} d\tau\int \Dp~ [ \dot y(p,\tau)\dot y(-p,\tau)+  (p^2+m^2) y(p,\tau)y(-p,\tau)]  }
\ee

\section{Two point Function} \label{4}

We will evaluate the boundary two point function through a bulk calculation. Since the bulk simply implements an Exact RG we are guaranteed to get the right answer. Nevertheless we outline the calculation for completeness.

We take $\tau_i=0~(i.e.~\lm_i =\infty)$ and $ \tau_f=\infty ~(i.e.~\lm_f=0)$. So $G_i=\frac{1}{p^2}$ and $G_f=0$.
We set $m=0$ for simplicity so that $\bp=p$.

{\bf In terms of $\phi$:}

We first consider the action with $\phi$ before the field redefinition.
We write
\[
\hf\int_p\int_{\tau_i}^{\tau_f} d\tau~ \frac{1}{\dot G(p)} \dot \phi(p)\dot \phi(-p)= \hf \int_p
\int_{G_i}^{G_f}	 dG(p)~(\frac{d\phi(p)}{dG(p)})(\frac{d\phi(-p)}{dG(p)}) 
\]
so that $G(p,\tau)$ is the RG-time for a mode $\phi(p)$.

The EOM is
\[
\frac{d^2\phi(p)}{dG^2}=0
\]
So 
the solution is
\[
\phi(p,G(p)) = b(p) G(p)+c(p)
\]
We impose boundary condition $\phi_f=\phi(p,\tau_f)=\phi(p,G_f)=0$ and $\phi(p,G_i)=\phi_i$.
This fixes $c=0$ and $b= \frac{\phi_i}{G_i}$. 
We add a perturbation $\int_p J(p)\phi_i(-p)$ at the boundary. The boundary term that comes from the variation of the action then obeys:
\be  \label{phii}
\phi_i(p) \frac{d\phi(-p,G)}{dG}|_{G=G_i}=J(-p)\phi_i(p)
\ee

So $b(p)=J(p)$ and our solution becomes
\[
\phi(p,G)= J(p) G(p)
\]
Plugging this solution into the action gives
\[
\hf \int_p \phi_i(p) \frac{d\phi(-p,G)}{dG}|_{G=G_i}= \hf \int_p J(p)J(-p) G_i(p)
=\hf\int_p  {J(p)J(-p)\over p^2}
\]
as expected.

{\bf In terms of $y$:}

The equation of motion is 
\[
-\frac{d^2y}{d\tau^2} + p^2 y=0
\]
\[
y(p,\tau)=a~ cosh (p\tau) + b~ sinh (p\tau)
\]
Boundary Condition:
\[
f_i y_i =\phi_i =JG_i,	~~f_fy_f=\phi_f=J G_f
\]
From \eqref{f1}, \eqref{G1} $a=-b=\frac{JA}{p}$
\[
f(p,\tau)= A(p)~ sech(p\tau)
\]
\[
y_f= J\frac{A}{p} ((sinh (p\tau_f)- cosh(p\tau_f)) \to 0 ,~~~\tau_f\to \infty
\]
\[
y_i=y(p,\tau =0) = \frac{JA}{p}
\]
We also have from \eqref{phii}
\[
y_i \ddx{y}{\tau}|_{\tau_i}= Jy_if_i \implies  \ddx{y}{\tau}|_{\tau_i}=Jf_i
\]

On shell action is
\be
\int_p \hf y_i(p) \ddx{y(-p)}{\tau}|_{\tau_i}=\int_p \hf \frac{J(p)A}{p} J(-p)f_i(p)= \hf\int_p \frac{J(p)A^2J(-p)}{p}=\hf\int_p \frac{J(p)J(-p)}{p^2}
\ee
where \eqref{A} has been used.

\section{Composite fields in $O(N)$ models} \label{5}

The $O(N)$ model at the free fixed point is 
\be
\sqrt N\sum_{I=1}^N\hf\int \Dp  \phi^I(p) p^2 \phi^I(-p)
\ee

 The composite operators considered in 
 \cite{Dharanipragada:2022} are
 \br
\si (x)&=& \sum _{I=1}^N \phi^I(x) \phi_I(x)\\
\si_\mu^{IJ} &=& \phi^I \olra{\p_\mu} \phi^J\\
\si_{\mu\nu}&=& \Theta_{\mu\nu}
\er
$\Theta_{\mu\nu}$ is the transverse and traceless (improved) energy momentum tensor.

The ERG equation for the action for these fields was worked out in \cite{Dharanipragada:2020} for the free part and the cubic interactions were worked out in \cite{Sathiapalan:2020,Dharanipragada:2023}.

\subsection{Kinetic term}

 The  free part of the evolution operator for the ERG equation for the fundamental scalar \eqref{evofunda}, for the composite scalar, vector and graviton, found in \cite{Dharanipragada:2020} are ($\mu =1,...,D$):

\be \label{evoscalar}
e^{-S_{\lm_f}[\si_{ f}]}=\int {\cal D}\si_{ i}(q)\int _{\si(q,\tau_i)=\si_{ i}(q)}^{\si (q,\tau_f)=\si_{ f}(q)}{\cal D}\si(q,\tau)e^{-\hf\int_{\tau_i}^{\tau_f} d\tau\int_q \frac{1}{ \dot I(q)} \dot \si(q)\dot \si(-q)}e^{-S_{\lm_i}[\si_{ i}]}.
\ee

\be \label{evovector}
e^{-S_{\lm_f}[\si_{\mu f}]}=\int {\cal D}\si_{\mu i}(q)\int _{\si_\mu(q,\tau_i)=\si_{\mu i}(q)}^{\si_\mu (q,\tau_f)=\si_{\mu f}(q)}{\cal D}\si_\mu^{IJ}(q,\tau)e^{-\hf\int_{\tau_i}^{\tau_f} d\tau\int_q \frac{1}{16 \dot I_v(q)} \dot \si_\mu ^{IJ}(q)\dot \si^{\mu IJ}(-q)}e^{-S_{\lm_i}[\si_{\mu i}]}.
\ee

\be   \label{evograv}
e^{-S_\lm[\si^{\mu\nu}_f]}=\int \CD \si^{\mu\nu}_i(q)\int _{\si^{\mu\nu}(q,\tau_i)=\si^{\mu\nu}_i(q)}^{\si^{\mu\nu}(q,\tau_f)=\si^{\mu\nu}_f(q)}\CD \si^{\mu\nu}(q,\tau)
e^{-\hf\int_{\tau_i}^{\tau_f}d\tau~\int_q \frac{\dot \si^{\mu\nu}(q,\tau)\dot \si_{\mu\nu}(q,\tau)}{\dot I_t(q)}}e^{-S_{\lo}[\si^{\mu\nu}_i]}.
\ee

$I(q)$ is the {\em regulated} one loop integral with {\em low energy} propagators for the fundamental scalars\footnote{Note that we will use the flipped ERG defined in \cite{Dharanipragada:2023} here.}: 
\[
I(q)=\int \Dp \frac{K(p+q)K(p)}{(p+q)^2p^2}
\]
$K(p,\lm)$ is a cutoff function. In ERG a standard choice is $K=e^{-\frac{p^2}{\lm^2}}$. Here we will specify (see below) the form of $I(q)$. This defines $K$ implicitly. We will not need the explicit form of $K$.

In position space $I(x)= \DD_l(x)^2$.
$I_v,I_t$ are the corresponding regulated propagators for the transverse vector and graviton respectively. When $\lm\to \infty$,
\[
I(q)\approx (q^2)^{\Dt-2},~~16I_v(q) \approx (q^2)^{\Dt-1},~~~I_t(q) \approx (q^2)^\Dt
\]
Transversality of the currents are assumed: $\p^\mu \si_\mu= \p^\mu \si_{\mu\nu}=0$. Also $\si^\mu_{~\mu}=0$.

Now that the map is in place one can very easily take the results of \cite{Sathiapalan:2020, Dharanipragada:2020} and obtain the bulk kinetic terms for the composite scalar, gauge field - dual to the conserved vector current, and graviton - dual to the energy momentum tensor - by applying the following maps to the actions given in \eqref{evoscalar},\eqref{evovector} and \eqref{evograv}.

\br 
\si(p) &=& f(p) y(p)\\
\si_\mu (p)&=&f(p)y_\mu(p)\\
\si_{\mu\nu}(p)&=&f(p) y_{\mu\nu}(p)
\er
where $f$ is given by \eqref{f1}. $A(p)$ in each case is fixed by \eqref{g1}. There is an issue involving a simultaneous map for all three fields that was discussed and resolved in \cite{Dharanipragada:2020,Dharanipragada:2023}. Those comments apply here also but will not be spelt out.

\begin{enumerate}
\item Scalar

In $D=3$, $\si(x)$ has dimension one. So in the UV limit 
\[
\langle \si(x) \si (0)\rangle =G_s(x) \approx \frac{1}{x^2}
\]
So
\be  \label{gs1}
G_s(p)= \frac{1}{p} 
\ee

We use \eqref{G1} in the limit $\lm \to \infty$ to fix $A(p)$.
 \be  \label{G2}
G_s(p,\lm)= -\frac{A(p)^2}{p} (tanh(\frac{p}{\lm})-1)
\ee
This requires $A(p)=1$ in order to obtain \eqref{gs1}.
Thus
\be  \label{G3}
G_s(p,\tau)= -\frac{1}{p} (tanh(p\tau)-1)
\ee

So from \eqref{f1}
\be  \label{f2}
f_s(p)=  ~sech ~(p\tau)
\ee
$f(p)\to 1$ in the UV limit.
\[
\si (p) = f(p) y(p) \approx y(p)
\]

Thus in \eqref{evoscalar} we set $I(p)=G_s(p)$.
\item

$\si_\mu(x)$ has dimension 2. So when $\lm =\infty$,
\[
G_v(p)=p \implies A(p)=p
\]
\be
G_v(p,\tau)=-p~(tanh(p\tau)-1),~~~f_v(p,\tau)= p ~sech(p\tau)
\ee

Thus we will set in \eqref{evovector}, $16 I_v(q)=G_v(q)$.

\[
\si_\mu (p)=p y_\mu(p) =G(p) y_\mu(p) \implies y_\mu(p)\approx A_\mu(p)
\]
where $A_\mu$ is the external background gauge field in the boundary theory.
So $y_\mu(p)$ should be identified with the source $A_\mu(p)$ at $\lm =\infty$.

\item
$\si_{\mu\nu}(x)$ has dimension 3. So similarly, 
\be   \label{tensor}
G_t(p) = p^3 \implies A(p)=p^2
\ee
\be  \label{G4}
G_t(p,\tau)= -p^3 (tanh(p\tau)-1)
\ee

Once again in \eqref{evograv} we set $I_t(q)=G_t(q)$.

\be  \label{f3}
f_t(p,\tau) = p^2 sech~p\tau
\ee
\be  \label{siy}
\si_{\mu\nu}(p) = p^2 y_{\mu\nu}(p)
\ee
This gives a (unusual) relation between the boundary value of the bulk metric perturbation $y_{\mu\nu}$ and the boundary energy momentum tensor.

\end{enumerate}

The $y_\mu ,y_{\mu\nu}$ will be identified with the gauge fixed transverse and traceless bulk fields.
Thus the bulk gauge field satisfies additionally the ``holographic" gauge condition $y_\tau=0$ and metric perturbation,  $y_{\tau \mu}=0=y_{\tau\tau}$.

These actions are mapped to
\be
S=\hf \int _0^\infty d\tau\int \Dp [\dot y(p)\dot y(-p) + p^2y(p)y(-p)]
\ee 
\be
S=\hf \int_0^\infty d\tau\int \Dp [\dot y_\mu(p)\dot y^\mu(-p) + p^2y^\mu(p)y_\mu(-p)]
\ee 
\be  \label{ymunu}
S=\hf \int_0^\infty d\tau\int \Dp [\dot y^{\mu\nu}(p)\dot y_{\mu\nu}(-p) + p^2y^{\mu\nu}(p)y_{\mu\nu}(-p)]
\ee 
respectively. We have chosen the parameter $m=0$ so that there is the possibility of gauge invariance - only then can these be interpreted as gauge fixed versions of the Euclidean Maxwell action or linearized Einstein-Hilbert action respectively.  When  $m\neq 0$ the fields would be massive and may be required for boundary operators that do not correspond to conserved currents. Note that $m$ does not correspond to any parameter in the CFT (unlike in the AdS case). It is a parameter in the regularization scheme used in the RG.

\subsection{Interactions}

\subsubsection{cubic scalar}

The cubic interaction term in the evolution operator for the ERG equation for the scalar field action was worked out in \cite{Sathiapalan:2020}.

\[ V(\sigma, \sigma_{\mu\nu})=\frac{1}{\sqrt N}\int\limits_{k_1,k_2,k_3}\dd(k_1+k_2+k_3)\times
\]
\[
\bigg\{-\frac{2}{3}\frac{d}{d\tau} \Big[\int_p\DD(p)\DD(p+k_1)\DD(p+k_1+k_2)\Big]_{regulated,\lm}\Big( \frac{\sigma(k_1,\tau)}{G_s(k_1,\tau)} \frac{\sigma(k_2,\tau)}{G_s(k_2,\tau)} \frac{\sigma(k_3,\tau)}{G_s(k_3,\tau)}\Big)\bigg\}
\]

We now map the action to flat  space by the substitution  $\sigma = fy$ and choosing $f$ judiciously. Thus the coefficient of $y(k_1)y(k_2)y(k_3)$ becomes
\[
\frac{1}{\sqrt N}\int\limits_{k_1,k_2,k_3}\dd(k_1+k_2+k_3)\times
\]
\be \label{sss1}
\bigg\{-\frac{2}{3}\frac{d}{d\tau} \Big[\int_p\Delta(p)\Delta(p+k_1)\Delta(p+k_1+k_2)\Big]_{regulated,\lm}\Big( \frac{f(k_1,\tau)}{G_s(k_1,\tau)} \frac{f(k_2,\tau)}{G_s(k_2,\tau)} \frac{f(k_3,\tau)}{G_s(k_3,\tau)}\Big)\bigg\}
\ee
The time derivative of the loop momentum integral with a convenient regularization procedure is calculated in \cite{Dharanipragada:2023} and is (some details are given in Appendix \ref{a1}),
\be \label{sss2}
 \frac{\tau^{-1+2\eps}}{\Gamma(\eps)}\times~2(\frac{k_1}{ \tau})^{-\nu_1}K_{\nu_1}(k_1\tau)\times
2(\frac{k_2}{ \tau})^{-\nu_2}K_{\nu_2}(k_2\tau)\times
2(\frac{k_3}{ \tau})^{-\nu_3}K_{\nu_3}(k_3\tau)
\ee
upto a normalising factor, and  with $\nu_i=\hf ~;~ i=1,2,3$. Here $\eps =D-3$.
For each $i$ we have
\be \label{num}
 (k_i\lm)^{-\hf} K_\hf(k_i/\lm)= \frac{e^{-k_i/\lm}}{k_i}
\ee
Using \eqref{f2} and \eqref{G2} we get
with $A=1$,
\be  \label{den}
\frac{f}{G}= \frac{k_i}{A(k)} \frac{sech (k_i\tau)}{(tanh~({k_i\tau})-1)}=-k_i e^{k_i/\lm}
\ee
All the $k$ dependence cancels in the product of \eqref{num} and \eqref{den}! So we get a local vertex here just as in the $AdS$ case.
\be  \label{cubics}
S[y]=\frac{1}{\Gamma(\eps)}\int d\tau \int_{k_1,k_2,k_3}\tau^{-1+2\eps} y(k_1,\tau)y(k_2,\tau)y(k_3,\tau)~\dd(k_1+k_2+k_3)
\ee

 However there is an important difference. Here the cancellation happens only when $D=3$ which is when $\nu =\hf$. In the AdS case it happens for any $D$. Note also the dependence on $\tau$ which is unusual. $y(x)$ has scaling dimension one (in $D=3$), so the factor of $\tau^{-1}$ makes for the scale invariance of the interaction.
  $\eps$ is a regulator $D=3+\eps$. The interaction seems to vanish in $D=3$. However the boundary correlation function generated by this interaction is finite - see below \cite{Sathiapalan:2020}.
  
The interaction term \eqref{cubics} is not general coordinate invariant in $D+1$ dimensions due to the factor of $\frac{1}{\tau}$. This factor is required for scale invariance. It can perhaps be understood as arising from a dilaton coupling $e^\Phi$ with 
  \[
\langle e^\Phi\rangle = \frac{1}{\tau}
\]   
 This would restore general coordinate invariance of the action. 
 
\subsubsection{Scalar Correlation function}
 The correlation function of three scalars in the boundary theory is
 
\[
\langle \langle \si(k_1)\si(k_2)\si(k_3)\rangle \rangle = \int \Dp \frac{1}{p^2 (p+k_1)^2(p+k_1+k_2)^2}
\]
\[
=
\frac{16}{\Gamma(\eps)}\int_0^\infty d\tau ~\tau^{-1+2\eps}(\frac{k_1}{ \tau})^{-\hf}K_{\hf}(k_1\tau)
(\frac{k_2}{ \tau})^{-\hf}K_{\hf}(k_2\tau)
(\frac{k_3}{ \tau})^{-\hf}K_{\hf}(k_3\tau)
\]
\be \label{sc2}
=
\frac{16~ \Gamma(2\eps)}{\Gamma(\eps)}(k_1+k_2+k_3)^{-2\eps}\frac{1}{k_1k_2k_3}
\ee
 When $\eps\to 0$ we get
\be  \label{sc3}
\langle \langle \si(k_1)\si(k_2)\si(k_3)\rangle \rangle\approx \frac{1}{k_1k_2k_3}
\ee

We will now recover  this from the bulk calculation in the semi classical approximation using the Witten diagrams involving the cubic vertex \eqref{cubics}. The external fields $y(k_i,\tau)$ are given by
\be  \label{sclass}
y(k_i,\tau)=\frac{G_s(k_i,\tau)}{f(k_i,\tau)} J(k_i)
\ee
Here $J(k_i)$ is a source for $\si(k_i)$ in the boundary. Using \eqref{G3} and \eqref{f2} we get
\[
y(k_i,\tau)= \frac{e^{-k_i\tau}}{k_i} J(k_i)
\]

Plugging this into the action \eqref{cubics} and doing the integral over $\tau$ one obtains (when $\eps \to 0$),
\be
W[J_i]\approx\int_{k_1,k_2,k_3}\dd(k_1+k_2+k_3)\frac{1}{k_1k_2k_3}J(k_1)J(k_2)J(k_3)
\ee
 and thus \eqref{sc3}.

\subsubsection{scalar-scalar-spin 2}
\label{ssg}

The interaction term in this case is 
(\cite{Dharanipragada:2023})

\be \label{v}
-4  \frac{d}{d\tau}\Big(\int_p(\DD_l (p+k_1+k_2)\DD_l(p+k_1) p^\mu p^\nu\DD_l(p))\Big)\Big( \frac{f_s(k_1,\tau)}{G_s(k_1,\tau)} \frac{f_s(k_2,\tau)}{G_s(k_2,\tau)} \frac{f_t(k_3,\tau)}{G_t(k_3,\tau)}\Big)\bigg\}.
\ee

For the  loop integral in this, the computation is done in an appendix in \cite{Dharanipragada:2023}, resulting in
\begin{equation} \label{v1}
	k_{1}^\mu k_{2}^\nu\times\tau^{2D-3}\Big(\frac{k_1}\tau\Big)^{-\nu}K_{\nu}(k_1\tau)~
	\Big(\frac{k_2}\tau\Big)^{-\nu}K_{\nu}(k_2\tau)~
	\Big(\frac{k_3}\tau\Big)^{\nu'}K_{\nu'}(k_3\tau),
\end{equation}
where  $\nu=2-\Dt=\hf $ and $\nu' =\frac{D}{2}=\frac{3}{2}$. We take $D=3$. We have already seen that $\frac{f_s}{G_s}$ cancels the $\nu=\hf$ terms. That leaves the tensor term. 
\be   \label{tensor1}
\frac{f_t}{G_t} = \frac{k_3}{A(k_3)}\frac{sech (k_3\tau)}{(tanh~({k_3\tau})-1)}=-\frac{1}{k_3} e^{k_3\tau}
\ee
We have used \eqref{tensor}. On the other hand for $\nu'=\frac{3}{2}$,
\be  \label{tensor2}
(\frac{k_3}\tau\Big)^{\nu'}K_{\nu'}(k_3\tau)\approx \frac{e^{-k_3\tau}}{\tau^3}(1+k_3\tau)
\ee 

Multiplying \eqref{tensor1} and \eqref{tensor2} we get $\frac{1}{k_3}(1+k_3\tau)$.
Thus the scalar-scalar-spin 2 coupling is non local:
\be   \label{spin2ss}
\hf\int d\tau~\int_{k_1,k_2,k_3}k_1^\mu y(k_1,\tau)k_2^\nu y(k_2,\tau)y_{\mu\nu}(k_3)\frac{1}{k_3}(1+k_3\tau) \dd(k_1+k_2+k_3)
\ee

The dimension of $y_{\mu\nu}(x)$ is one (in $D=3$) and one can see that the interaction term is scale invariant.

\subsubsection{Scalar-scalar-spin 2 correlation}

The boundary correlation function is given by (for the transverse traceless field)
\[
\langle \langle \si(k_1)\si(k_2)\si_{\mu\nu}(k_3)\rangle \rangle \approx \int \Dp \frac{p_\mu p_\nu}{p^2 (p+ k_1)^2(p+k_1+k_2)^2}
\]
\[
=k_{1\mu}k_{2\nu} \int d\tau ~
\tau^{2D-3}\Big(\frac{k_1}\tau\Big)^{-\hf}K_{\hf}(k_1\tau)~
	\Big(\frac{k_2}\tau\Big)^{-\hf}K_{\hf}(k_2\tau)~
	\Big(\frac{k_3}\tau\Big)^{\frac{3}{2}}K_{\frac{3}{2}}(k_3\tau)
\]
\be  \label{corspin2}
\approx k_{1\mu}k_{2\nu} \frac{k_1+k_2+2k_3}{k_1k_2(k_1+k_2+k_3)^2}
\ee

We now recover this from a bulk calculation using the interaction term
\eqref{spin2ss}.

Plug in the classical solution for $y$ \eqref{sclass} and (using \eqref{G4}, \eqref{f3})
\be  \label{tclass}
 y_{\mu\nu}(k_3,\tau)=\frac{G_t(k_3,\tau)}{f_t(k_3,\tau)}J_{\mu\nu}(k_3)=k_3e^{-k_3\tau}J_{\mu\nu}(k_3)
\ee
where $J_{\mu\nu}$ is a source for the boundary field $\sigma_{\mu\nu}$ (i.e. energy momentum tensor), 
into \eqref{spin2ss}, we get on doing the $\tau$ integral
\be
W[J,J_{\mu\nu}]\approx\int_{k_1,k_2,k_3}\dd(k_1+k_2+k_3) k_{1\mu} k_{2\nu}\frac{(k_1+k_2+2k_3)}{k_1k_2(k_1+k_2+k_3)^2} J(k_1)J(k_2)J^{\mu\nu}(k_3)
\ee

which reproduces \eqref{corspin2}.

That concludes our discussion of the spin 2  and scalar correlation functions.

Regarding higher spins: we also do not expect an exact cancellation in all the higher spin (spin greater than 2)  scalar interactions  where also $K_\nu$ with $\nu >\hf$ are involved and thus they will also be non local.

\subsection{General Coordinate Invariance}

\subsubsection{Kinetic term for graviton}

Consistency of massless spin 2 theories requires gauge invariance to get rid of the negative norm states. Linearized gauge invariance takes the form: ($A$ runs over $\tau$ and the $D$ boundary directions $\alpha$)
\be  \label{gt}
\dd h_{MN} = \p_{(M}\eps_{N)}
\ee

The quadratic part of the Einstein-Hilbert action with the above invariance is
\be  \label{haction}
S=-\frac{1}{4} h_{AB,M}h^{AB,M} +
\frac{1}{4} h^A_{~A,M}h^{B~M}_{~B,}
-\frac{1}{2} h^A_{~A,B}h^{B~M}_{~M,}
+\frac{1}{2} h^{M~A}_{~B,}h^{B}_{~A,M}
\ee
One can use this gauge invarance to choose 
the usual ``holographic" gauge
\be   \label{hg}
h_{\tau \alpha}=h_{\tau\tau}=0
\ee

The equations of motion are:
\be
\DDx{S}{h_{AB}}= \hf h^{AB,M}_{~~~~M}-\hf \dd^{AB}h^{C~~M}_{~C,~M}+\hf \dd^{AB}h^{C~~M}_{~M,~C}
+\hf h^{C~~AB}_{~C,}-\hf h^{MA,B}_{~~~~~M}
-\hf h^{MB,A}_{~~~~~M}=0
\ee

Separating the $\tau$ index and imposing the holographic gauge gives the following three equations:
\be  \label{ab}
\DDx{S}{h_{\alpha\beta}}= \hf h^{\alpha\beta,M}_{~~~~M}-\hf \dd^{\alpha\beta}h^{\gamma~~M}_{~\gamma,~M}+\hf \dd^{\alpha\beta}h^{\gamma~~\mu}_{~\mu,~\gamma}
+\hf h^{\gamma~~\alpha\beta}_{~\gamma,}-\hf h^{\mu\alpha,\beta}_{~~~~~\mu}
-\hf h^{\mu\beta,\alpha}_{~~~~~\mu}=0
\ee
\be  \label{tb}
\DDx{S}{h_{\tau\beta}}=\hf h^{\gamma~~\tau\beta}_{~\gamma,}-\hf h^{\mu \beta,\tau}_{~~~~~\mu}=0
\ee
\be   \label{tt}
\DDx{S}{h_{\tau\tau}}=-\hf h^{\gamma~~\mu}_{~\gamma,~\mu} +\hf h^{\gamma~~\mu}_{~\mu,~\gamma}=0
\ee
Taking the trace of \eqref{ab} gives:
\[
\hf(2-D) h^{\alpha~~\mu}_{~\alpha,~\mu}+\hf (1-D)h^{\alpha~~\tau}_{~\alpha,~\tau}
+\hf(D-2)h^{\gamma~~\mu}_{~\mu,~\gamma}=0
\]
Using \eqref{tt} this simplifies to
\be
\hf (1-D) h^{\alpha~~\tau}_{~\alpha,~\tau}=0
\ee
This has only the solution
\be
h^\alpha_{~\alpha}=c_0(p)+c_1(p) \tau
\ee
where $p$ is the momentum in the boundary directions. Since this couples to the trace of the boundary energy momentum tensor, which is zero, $h^\alpha_{~\alpha}$ can be set 
to zero at $\tau=0$ and also at $\tau=\infty$, which sets $c_0,c_1$ to zero. But one is not forced to do so. We leave it as zero for the moment.

\eqref{tb} then gives the constraint
\be
\ppx{}{\tau} h^{\mu\beta}_{~~,\mu}=0
\ee

If $h^{\mu\beta}_{~~,\mu}$ vanishes at $\tau=0$ it is zero everywhere. In the gauge \eqref{hg} we have a remaining gauge invariance
\be
\dd h_{\alpha \beta} = \p_{(\alpha}\eps_{\beta)}
\ee
where $\eps$ is independent of $\tau$. This can be used to set $h^{\mu\beta}_{~~,\mu}=0$ on one slice. So it is zero everywhere. Thus we have a transverse and traceless metric perturbation in $D+1$ dimensional flat space with a boundary at $\tau=0$. And the final equation for it is
\be
h_{\alpha \beta,~~M}^{~~~~M}=0
\ee

\subsubsection{Non locality of spin 2 coupling to Scalar}

The interaction term \eqref{spin2ss}
\be   \label{spin2ss1}
\hf\int d\tau~\int_{k_1,k_2,k_3}k_1^\mu y(k_1,\tau)k_2^\nu y(k_2,\tau)y_{\mu\nu}(k_3)\frac{1}{k_3}(1+k_3\tau) \dd(k_1+k_2+k_3)
\ee
is non local because of the $\frac{1}{k_3}$ factor. This suggests a field redefinition of the spin2 field:
\be  \label{redef}
y_{\mu\nu}(k,\tau)=k h^b_{\mu\nu}(k,\tau)
\ee
With this redefinition \eqref{siy}
gives
\be
\si_{\mu\nu}(\tau,p)=p^3 h_{\mu\nu}^b(\tau,p)
\ee
Since the Green function for the energy momentum tensor in the boundary theory is also $\approx p^3$ we see that the bulk field $h_{\mu\nu}^b$ at the boundary, $h_{\mu\nu}^b (\tau=0)$, can be identified with the boundary background metric  - the source ($J_{\mu\nu}$) for the energy momentum tensor  - (which was not possible with $y_{\mu\nu}$).  

Then
\eqref{ymunu} becomes
\be  \label{hmunu}
S=\hf \int_0^\infty d\tau\int \Dp p^2[\dot h^{b\mu\nu}(p)\dot h^b_{\mu\nu}(-p) + p^2h^{b\mu\nu}(p)h^b_{\mu\nu}(-p)]
\ee 

and \eqref{spin2ss1} becomes

\be   \label{spin2ss2}
\hf\int d\tau~\int_{k_1,k_2,k_3}k_1^\mu y(k_1,\tau)k_2^\nu y(k_2,\tau)h^b_{\mu\nu}(k_3)(1+k_3\tau) \dd(k_1+k_2+k_3)
\ee

$h^b_{\mu\nu}(x)$ is dimensionless as is usual in descriptions of the metric perturbation: $g_{\mu\nu}= \dd_{\mu\nu}+\kappa h^b_{\mu\nu}$, where $\kappa$ has dimensions of length. \eqref{hmunu} is dimensionless and does not require the dimensionful Newton's constant $G_N=\kappa^2$. In the AdS case one would have had ${\kappa\over R_{AdS}} \approx \frac{1}{\sqrt N}$. In the present case, in flat space, there is no length scale corresponding to $R_{AdS}$.

\eqref{haction} has the linearized gauge invariance in $D+1$ dimensions given by \eqref{gt}.

Clearly if we multiply the entire action by $p^\mu p_\mu$ it continues to have the invariance. Thus we can start with the four derivative action for $h^b_{MN}$ given in momentum space by

\[
S=\int_p p^\mu p_\mu\Big[-\frac{1}{4} p^Mp_M h^b_{AB}(p)h^{bAB}(-p) +
\frac{1}{4} p_Mp^Mh^{bA}_{~A}(p)h^{bB~}_{~B}(-p)
\]
\be  \label{hBactionmom}
-\frac{1}{2} p_Bp^M h^{bA}_{~A}(p)h^{bB}_{~M}(-p)
+\frac{1}{2}p^Ap_M h^{bM~}_{~B}(p)h^{bB}_{~A}(-p)\Big]
\ee
and it is invariant under
\be \label{gt6}
\dd h^b_{MN}(p)= ip_{(M}\tilde \eps_{N)}
\ee
Gauge fixing it gives \eqref{hmunu}. The reason for the tilde will become clear below.

\eqref{hmunu} is written partially in position space and partially in momentum space. We let $M=\mu$ be the directions along the boundary and $M=\tau$ be the bulk direction. The invariance,\eqref{gt}, of \eqref{haction} in this notation is:

\br 
\dd h_{\mu\nu}(\tau,p)&=& ip_{(\mu}\eps_{\nu)}(p,\tau) \label{gt1}\\
\dd h_{\mu\tau}(\tau,p)&=& ip_{\mu}\eps_{\tau}(\tau,p)+ \ddx{}{\tau}\eps_\mu(\tau,p) \label{gt2}\\
\dd h_{\tau\tau}(\tau,p)&=& 2\ddx{}{\tau}\eps_{\tau}(\tau,p) \label{gt3}\\\nonumber
\er

\subsubsection{Gauge invariance of scalar coupling}
The minimal linearized coupling of the graviton perturbation to a scalar field is given by
\be \label{minc}
S_I=\hf \int _x [\p^M \phi \p_M \phi(1+ \hf \dd^{AB} h_{AB})+
\p_M\phi \p_N\phi ~h^{MN}]
\ee
and is invariant under
\br
\dd \phi &=& \eps ^M \p_M\phi\\
\dd h_{MN} &=& \p_{(M}\eps_{N)}\\\nonumber
\er

We define $g^{MN}=\dd^{MN}+h^{MN}$ so that
to linear order $h_{MN}=-h^{MN}$.

 Writing $M=(\tau,\mu)$ we get 
\[
S_I= \hf \int d\tau \int_p  [\dot \phi(\tau,p) \dot \phi(\tau,-p) + p^\mu p_\mu \phi (\tau,p)\phi(\tau,-p)]
\]
\[+ \hf \int d\tau \int_{p_1,p_2,p_3}  [\dot \phi(\tau,p_1) \dot \phi(\tau,p_2) - p_1^\mu p_{2\mu} \phi (\tau,p_1)\phi(\tau,p_2)]\hf(h^\tau_{~\tau}(\tau,p_3)+h^\mu_{~\mu}(\tau,p_3))\dd(p_1+p_2+p_3)
\]
\[
+ \hf \int d\tau \int_{p_1,p_2,p_3} [\dot \phi(\tau,p_1) \dot \phi(\tau,p_2) h_{\tau\tau}(\tau,p_3)- p_{1\mu}  \phi (\tau,_1)p_{2\nu}\phi(\tau,p_2) h^{\mu\nu}(\tau,p_3)
\]
\be 
+2i\dot \phi(\tau,p_1)p_{2\mu} \phi(\tau,p_2)h^{\tau\mu}(\tau,p_3)]\dd(p_1+p_2+p_3)
\ee

The invariance is \eqref{gt1}, \eqref{gt2},\eqref{gt3} and
\be
\dd \phi = \eps ^\tau \dot \phi(\tau,p)  +\eps ^\mu ip_\mu \phi (\tau,p)
\ee

Gauge fixing to the holographic gauge and using the tracelessness property we get
\[
S_I = \hf \int d\tau \int_p  [\dot \phi(\tau,p) \dot \phi(\tau,-p) + p^\mu p_\mu \phi (\tau,p)\phi(\tau,-p)]
\]
\be   \label{SI}
-\hf \int d\tau \int_{p_1,p_2,p_3}  p_{1\mu}  \phi (\tau,p_1)p_{2\nu}\phi(\tau,p_2) h^{\mu\nu}(\tau,p_3)\dd(p_1+p_2+p_3)
\ee

Now consider the interaction term \eqref{spin2ss2}. It can be identified with the gauge fixed $S_I$ in \eqref{SI} if we let
\be  \label{map}
h_{\mu\nu}(\tau,p)=(1+p\tau)h_{\mu\nu}^b(\tau,p)
\ee

The question is can this be extended to a gauge invariant interaction with the scalar. It turns out to be possible and the gauge invariant coupling is given below (Note that the trace of the metric is chosen zero due to conformal invariance of the boundary theory):

\[
S_{II}= \int d\tau \int_{p_1,p_2,p_3}\dd(p_1+p_2+p_3)
\Big\{-\hf p_{1\mu} \phi(\tau,p_1) p_{2\nu} \phi(\tau,p_2)\underbrace{(1+p_3\tau)h^{b\mu\nu}(\tau,p_3)}_{h^{\mu\nu}}
\]
\[+i p_{1\mu} \phi(\tau,p_1)\p_\tau \phi(\tau,p_2)\underbrace{\Big[(1+p_3\tau)h^{b\mu\tau}(\tau,p_3)- \frac{ip_{3\nu}}{p_3}h^{b\mu\nu} (\tau,p_3)\Big]}_{h^{\mu\tau}}
\]
\be  \label{giact}
+ \hf \p_\tau \phi(\tau,p_1) \p_\tau \phi(\tau,p_1)\underbrace{\Big[(1+p_3\tau) h^{b\tau\tau} - 2 \frac{ip_{3\mu}}{p_3}h^{b\mu\tau}\Big]}_{h^{\tau\tau}}
\Big\}
\ee
Gauge invariance of this action requires 
\[
\dd h_{\mu\nu}= ip_{(\mu}\eps_{\nu)}=(1+p\tau)\dd h_{\mu\nu}^b
\]
Thus 
\be
\dd h_{\mu\nu}^b(\tau,p) = ip_{(\mu} \eps_{\nu)}(\tau,p)\frac{1}{(1+p \tau)}\equiv ip_{(\mu}\tilde\eps_{\nu)}(\tau,p)
\ee
with
\be   \label{gmap0}
\eps_\mu(\tau,p)=(1+p\tau)\tilde\eps _\mu(\tau,p) 
\ee

Similarly we relate $\tilde \eps _\tau$ and $\eps _\tau$:
\be  \label{gmap2}
 \eps_\tau =(1+p\tau)\tilde \eps_\tau (\tau,p)- \frac{ip^\mu \tilde\eps_\mu}{p}
\ee

We let
\br  \label{gt5}
\dd h_{\mu\nu}^b (\tau,p)
 &=& ip_{(\mu}\tilde \eps_{\nu)}\nonumber\\
\dd h_{\mu\tau}^b &=& ip_\mu \tilde \eps_\tau + \p_\tau \tilde \eps_\mu \nonumber\\
\dd h_{\tau\tau}^b &=& 2 \p_\tau \tilde \eps_\tau 
\er

Further if we let 
\br
h_{\mu\nu}(\tau,p)&=& (1+p\tau) h_{\mu\nu}^b(\tau,p)\label{fmap1}\\
h_{\mu\tau}(\tau,p)&=& (1+p\tau)h_{\mu\tau}^b(\tau,p)- \frac{ip^\nu}{p}h_{\mu\nu}^b (\tau,p)\label{fmap2}\\
h_{\tau\tau}(\tau,p)&=&(1+p\tau) h_{\tau\tau}^b(\tau,p) - 2 \frac{ip^\mu}{p}h_{\mu\tau}^b(\tau,p)\label{fmap3}
\er

one finds that 
\br
\dd h_{\mu\nu}(\tau,p) &=& ip_{(\mu}\eps_{\nu)}(\tau,p) \label{map1}\\
\dd h_{\mu\tau}(\tau,p) &=& ip_\mu \eps_\tau(\tau,p) +\p_\tau \eps _\mu(\tau,p)\label{map2}\\
\dd h_{\tau\tau}(\tau,p) &=& 2\p_\tau \eps_\tau(\tau,p)\label{map3}
\er

We further require 
\be
\dd \phi=\eps ^\tau \p_\tau \phi + \eps^\mu \p_\mu \phi
\ee
and we see that \eqref{giact} is the standard gauge invariant minimal coupling with $h_{MN}$.  The bulk field $h_{MN}^b$ obtained from ERG is related to it as indicated above. 

Using \eqref{gt5} we can set 
\[
(1+p\tau)h_{\mu\tau}^b = \frac{ip^\nu}{p}h_{\mu\nu}^b(\tau,p)
\]
to ensure the holographic gauge $h_{\mu\tau}=0$. Similarly
\[
(1+p\tau)h_{\tau\tau}^b = \frac{2ip^\nu}{p}h_{\nu\tau}^b(\tau,p)
\]
sets $h_{\tau\tau}=0$.

After all this is done we can see that the gauge fixed $S_I$ \eqref{spin2ss2} can be obtained from the gauge invariant $S_{II}$ with $h_{MN}$ replaced by $h^b_{MN}$ given by 
\eqref{fmap1},\eqref{fmap2} and \eqref{fmap3}.

Thus we conclude that the interaction term can be obtained from a gauge invariant term by gauge fixing.

The gauge transformation of the bulk ERG field $h_{MN}^b$ also has a standard gauge transformation law \eqref{gt3} and so the kinetic term obtained through ERG \eqref{hmunu} is also obtainable by gauge fixing from a standard gauge invariant kinetic term \eqref{hBactionmom}.

The linearized gauge invariance ensures that at the linearized level the action we obtain is consistent with general coordinate invariance.

\subsection{Locality}

Even after the redefinition \eqref{redef}, the interaction term \eqref{spin2ss2} has a nonlocality due to the factor $(1+p\tau)h^b_{\mu\nu}(p,\tau)$. $p= \sqrt {p^\mu p_\mu}$ is non analytic at $p^\mu=0$. However it is possible to rewrite this term as a local one {\em on shell}. The classical solution \eqref{tclass} in terms of $h_{\mu\nu}^b$ is
\be  \label{tclass1}
 h^b_{\mu\nu}(k_3,\tau)=\frac{G_t(k_3,\tau)}{f_t(k_3,\tau)}J_{\mu\nu}(k_3)=k_3^2e^{-k_3\tau}J_{\mu\nu}(k_3)
\ee
Then 
\[
(1+\tau k_3) h^b_{\mu\nu}(k_3,\tau)=(1-\tau\frac{\p}{\p \tau} )h^b_{\mu\nu}(k_3,\tau)
\]

and the interaction term \eqref{spin2ss2} becomes
\be   \label{spin2ss4}
\hf\int d\tau~\int_{k_1,k_2,k_3}k_1^\mu y(k_1,\tau)k_2^\nu y(k_2,\tau)(1-\tau\frac{\p}{\p \tau} )h^b_{\mu\nu}(k_3) \dd(k_1+k_2+k_3)
\ee
This is local.  The derivation of the interaction term in the ERG evolution operator used the semi classical expansion in an essential way and  the correlation function calculation here (and in the usual AdS/CFT computations) also use the classical solution. So the replacement \eqref{spin2ss4} using the classical solution \eqref{tclass1} does not involve a loss of generality.\footnote{A very similar procedure was used recently in \cite{Dharanipragada:2025} to show that the bulk interaction of gauge fields has a local gauge invariant Yang-Mills form.} Thus we conclude that the bulk action written in terms of $h_{\mu\nu}^b$ is local.

 We replace \eqref{gmap0} by:
\be   \label{ngmap1}
\eps_\mu(\tau,p)=(1-\tau\frac{\p}{\p \tau})\tilde\eps _\mu(\tau,p) 
\ee
\eqref{fmap1},\eqref{fmap2} and \eqref{fmap3} are replaced by
\br
h_{\mu\nu}(\tau,p)&=& (1-\tau\frac{\p}{\p \tau}) h_{\mu\nu}^b(\tau,p)\label{nfmap1}\\
h_{\mu\tau}(\tau,p)&=& (1-\tau\frac{\p}{\p \tau})h_{\mu\tau}^b(\tau,p)+ \frac{ip^\nu}{p^2}\p_\tau h_{\mu\nu}^b (\tau,p)\label{nfmap2}\\
h_{\tau\tau}(\tau,p)&=&(1-\tau\frac{\p}{\p \tau}) h_{\tau\tau}^b(\tau,p) + 2 \frac{ip^\mu}{p^2}\p_\tau h_{\mu\tau}^b(\tau,p)\label{nfmap3}
\er

and \eqref{gmap2} is replaced by
\be  \label{ngmap2}
 \eps_\tau =(1-\tau\frac{\p}{\p \tau})\tilde \eps_\tau (\tau,p)+ \frac{ip^\mu \p_\tau \tilde\eps_\mu}{p^2}
\ee

Finally, the gauge choices $h_{\mu\tau}=h_{\tau \tau}=0$ imply that
\[
(1-\tau\frac{\p}{\p \tau})h_{\mu\tau}^b(\tau,p)+ \frac{ip^\nu}{p^2}\p_\tau h_{\mu\nu}^b (\tau,p)=0
\]
\[
(1-\tau\frac{\p}{\p \tau}) h_{\tau\tau}^b(\tau,p) + 2 \frac{ip^\mu}{p^2}\p_\tau h_{\mu\tau}^b(\tau,p)=0
\]

The price paid for this locality is that the spin-2 kinetic term becomes quartic in derivatives.
One can  ask whether the field redefinition introducing $h_{\mu\nu}^b$ is essential and whether  the $\frac{1}{k}$ non locality in \eqref{spin2ss1} is really unavoidable without this. One may legitimately suspect that some more general procedure would allow us to obtain a local interaction {\em and} a kinetic term quadratic in derivatives. In the present ERG scheme we have at our disposal only the functions $f$ and $G$. These are determined by requiring that the kinetic term have the correct form: The time derivative term in the kinetic term fixes the relation between $G$ and $f$. The spatial derivative term then fixes the precise form of $f$. Thus there is no freedom to change the interacting term and this non locality is unavoidable. Furthermore this cubic vertex produces the correct boundary correlator and the factor $\frac{1}{k}$ was necessary for this. It is hard to imagine replacing it with something else with the same dimension and obtaining the same correct result for the boundary correlator. The field redefinition introducing $h_{\mu\nu}^b$ seems at present to be the only way to get rid of this non locality. In any case we leave this as an open question.

Whether these bulk theories are physically acceptable is not clear at this point. The fact that they are dual in the large $N$ limit, to a free scalar field theory in the boundary, suggests that they should be acceptable.

\section{Conclusion} \label{6}

In this paper we described an attempt to use the same method that was used in \cite{Sathiapalan:2017,Sathiapalan:2019} to map the ERG equation to a holographic RG equation in AdS, to obtain a bulk space that is flat. As in the AdS case our starting point is the ERG equation for the $O(N)$ model at the Gaussian fixed point. The bulk fields dual to the composite scalar, spin 1 current and the energy momentum tensor are a scalar, a vector and the spin 2 graviton. Local kinetic terms (in flat space) for these fields were obtained by this method. The vector action is the gauge fixed Maxwell action. Similarly the spin 2 field kinetic term is the gauge fixed Einstein-Hilbert action. Two point functions are calculated and give the expected answers. 

The cubic scalar self interaction and scalar-scalar-spin 2 interaction were also calculated using this method. The scalar self interaction turns out to be local - but only in $D=3$. This is unlike the AdS case  where it was local for all $D$ \cite{Dharanipragada:2023}. There is also a dependence on $\tau$ - the distance from the boundary. This can possibly be interpreted as being due to a background dilaton field. This would restore general coordinate invariance. The spin 2 interaction is found to be non local due to a factor $\frac{1}{p}$ - again unlike the AdS case where it was local \cite{Dharanipragada:2023}. A field redefinition renders the locality milder. This milder non locality can also be removed and a local interaction can be written (see \eqref{spin2ss4}) that is equivalent on shell and gives the same correlation function.
 After the redefinition the graviton kinetic term has four derivatives - reminiscent of conformal gravity. 

The cubic correlators are calculated from the bulk using the semiclassical approximation and found to agree with boundary results.
Of course, the correlators calculated are guaranteed to be correct because using this bulk action is equivalent to doing exact RG calculation.

 There is a qualitative difference with AdS - the mass of the bulk scalar field is unrelated to the scaling dimension of the boundary operator. In fact it is not a parameter in the CFT - it arises from the cutoff function - so it is a parameter in the RG scheme.
 
The symmetries of the bulk theory need to be investigated. The bulk action is scale invariant and has the Poincare symmetry of the boundary. The  scalar-scalar graviton interaction term is a gauge fixed version of some manifestly general coordinate invariant terms. Thus it has the linearized gauge invariance required for consistency with general coordinate invariance. However the bulk spin 2 field is not quite the usual metric perturbation. They are related by a field redefinition.

Whether these theories are physically acceptable is not clear. But they are dual to a free theory in the boundary which suggests that they should be acceptable. 

The method described here is farly general and can be used to construct  holographic duals of known boundary theories. There is a lot of freedom and one can obtain different bulk theories in different spaces for the same boundary theory. But what we learn is that generically they result in non local bulk interaction terms or, after some field redefinitions, non standard but local kinetic and interaction terms. Thus one fact that emerges is that $AdS$ spaces are special in that they seem to naturally give local interactions in the bulk. 
 
Finally it would be interesting to see whether this has any connection with other approaches  \cite{Strominger:2017,Bagchi:2016,Gupta:2020,Jain:2023}.
 These  approaches are designed for Minkowski space time - so a direct comparison is difficult.  Another possibility is to consider the analytically continued ERG equation giving evolution of a time slice in real time - this would be a holography for Minkowski space time. This approach has been used for holographic dS/CFT correspondence [\cite{Witten:2001}-\cite{Larsen:2002}] and in the ERG context was discussed in \cite{Dharanipragada:2022}. 
 
 {\bf Acknowledgements:} We would like to thank Ghanashyam Date and Nemani Suryanarayana for useful discussions.

\begin{appendices}
\renewcommand{\theequation}{\Alph{section}.\arabic{equation}}
\section{Cubic Vertex Integral}
\label{a1}

We give below some of the intermediate steps that lead to \eqref{sss2} and \eqref{v1}. An outline is given here - the full details of the calculation of integrals of this form are give in \cite{Dharanipragada:2023}.

The loop integral in \eqref{sss1} involves a regulated propagator. However it can be easily seen \cite{Dharanipragada:2023} that when the external fields are on shell they satisfy $\sigma(k_i,\tau)=G_s(k_i,\tau) J(k_i)$, and the factors multipying the loop momentum integral  have no $\tau$ dependence and the vertex is thus a total $\tau$ derivative. So the correlator depends only on the bare propagator (at $\tau=0$) and is thus independent of the regularization scheme. So we are free to use any regularization scheme to define a vertex.

We start with the unregulated Schwinger parametrization of the propagator and this results in an integral of the general form:

\be \label{gen-s}
I=\int ds_1\int ds_2\int ds_3 \frac{s_1^{a_1-1}s_2^{a_2-1}s_3^{a_3-1}}{(s_1+s_2+s_3)^{\frac{D}{2}+m}}
e^{-\frac{k_1^2s_2s_3+k_3^2s_1s_2+k_2^2s_1s_3}{(s_1+s_2+s_3)}}
\ee
Some change of variables 
\[
s_1=\alpha_1 t,~~s_2=\alpha_2 t,~~s_3=\alpha_3 t,~~~~s_1+s_2+s_3=t,~~~\alpha_1+\alpha_2+\alpha_3=1
\]
and : 
\be   \label{cov2}
\alpha_1 \alpha_2t=\beta_3,~~\alpha_1 \alpha_3t=\beta_2,~~\alpha_3 \alpha_2t=\beta_1
\ee
gives 
\[I=
\int d\beta_1~d\beta_2~d\beta_3
\Big(\frac{J}{\beta_1\beta_2\beta_3}\Big)^{-D-2m+a_2}\beta_1^{-\Dt-m+a_t-a_1-1}\beta_2^{-\Dt-m+a_t-a_2-1}\beta_3^{-\Dt-m+a_t-a_3-1}
\]
\be
\times 
e^{-k_1^2\beta_1-k_2^2\beta_2
-k_3^2\beta_3}
\ee

This can be written as:

\[
I= \frac{1}{4^{D+2m-a_t}\Gamma(D+2m-a_t)}\int_0^\infty dx~x^{D+2m-a_t-1}\int d\beta_1 ~\beta_1^{\underbrace{-\Dt-m+a_t-a_1}_{\nu_1}-1}e^{-k_1^2\beta_1-\frac{x}{4\beta_1}}
\]
\[
\int d\beta_2 ~\beta_2^{\underbrace{-\Dt-m+a_t-a_2}_{\nu_2}-1}e^{-k_2^2\beta_2-\frac{x}{4\beta_2}}\int d\beta_3 ~\beta_3^{\underbrace{-\Dt-m+a_t-a_3}_{\nu_3}-1}e^{-k_3^2\beta_3 -\frac{x}{4\beta_3}}
\]

Here $a_t=a_1+a_2+a_3$.

Now use
\be
\int_0^\infty d\beta~
 \beta^{\nu-1}e^{-k^2\beta -\frac{x}{4\beta}}=2^{1-\nu}(\frac{k}{\sqrt x})^{-\nu}K_\nu(k\sqrt x)
\ee
to obtain
\[
I=\frac{4^{-D-2m+a_t}}{\Gamma(D+2m-a_t)}
\int^\infty_{\frac{1}{\lm^2}} dx~x^{D+2m-a_t-1}
2^{1-\nu_1}(\frac{k_1}{\sqrt x})^{-\nu_1}K_{\nu_1}(k_1\sqrt x)2^{1-\nu_2}(\frac{k_2}{\sqrt x})^{-\nu_2}K_{\nu_2}(k_2\sqrt x)
\]
\be
\times 2^{1-\nu_3}(\frac{k_3}{\sqrt x})^{-\nu_3}K_{\nu_3}(k_3\sqrt x)
\ee
At this last step we have introduced a regulator $\lm$ which will be identified with $\frac{1}{\tau}$.
The $\tau$ derivative is easy to evaluate in this form and gives the results \eqref{sss2},\eqref{v} for appropriate values of $m,a_i$ - more details are given in \cite{Dharanipragada:2023}.  For instance \eqref{sss2} is obtained with $a_1=a_2=a_3=1$ and $m=0$ and $\nu_i=\hf$. \eqref{v1} requires $a_1=a_2=1, a_3=3$ amd $m=2$. Then $\nu_1=\nu_2=\hf$ and $\nu_3=-\frac{3}{2}=-\nu'$.

\end{appendices}

\end{document}